\documentclass[prd,showpacs,showkeys,preprint,nofootinbib]{revtex4}
\usepackage{color}
\usepackage{amsmath}
\usepackage{amssymb}
\usepackage{yfonts}[1988/10/03]
\usepackage{graphicx}
\usepackage{subfig}
\usepackage{float}
\usepackage{xcolor}
\newcommand {\beq}{\begin{equation}}
\newcommand {\eeq}{\end{equation}}
\newcommand {\beqa}{\begin{eqnarray}}
\newcommand {\eeqa}{\end{eqnarray}}

\begin{document}{\textsf{\today} 
\title{ Testing loop quantum gravity  by quasi-periodic oscillations: Rotating blackholes}

\author{ Jafar Khodagholizadeh} 
\affiliation{ Department of Physics Education, Farhangian University, P.O. Box 14665-889, Tehran, Iran.}
\author{ Ghadir Jafari} 
\affiliation{ Department of Physics Education, Farhangian University, P.O. Box 14665-889, Tehran, Iran.}\author{ Alireza Allahyari} 
\affiliation{ Department of Astronomy and High Energy Physics,
Faculty of Physics, Kharazmi University, P. O. Box 15614, Tehran, Iran.
}\author{Ali Vahedi} 
\affiliation{ Department of Astronomy and High Energy Physics,
Faculty of Physics, Kharazmi University, P. O. Box 15614, Tehran, Iran.
}
\begin{abstract}
 We investigate a compelling model of a rotating black hole that is deformed by the effects of loop quantum gravity (LQG). We present a simplified metric and explore two distinct geometries: one in which the masses of the black hole and white hole are equal, and another in which they differ. Our analysis yields the radius of the innermost stable circular orbits (ISCO), as well as the energy and angular momentum of a particle within this framework.
Additionally, we find the frequency of the first-order resonance separately. 
We constrain the model by the quasi-periodic oscillations (QPO)
of the X-ray binary GRO J1655-40. We show that 
   $\lambda=0.15^{+0.23}_{-0.14}$ at  $1\sigma$ confidence level for equal mass black hole and white hole geometry. For the other geometry we get $\lambda=0.11^{+0.07}_{-0.07}$ at $1\sigma$ confidence level.We encounter a degeneracy in the parameter space that hinders our ability to constrain $\lambda$ with greater precision.
\end{abstract}
\maketitle
\section{Introduction}
With the direct observation of the black hole $M87^{\star}$ by the Event Horizon Telescope (EHT), a large baseline interferometry array operating at a wavelength of $1.3~ mm$, the significance of describing such phenomena has grown. This telescope successfully imaged the horizon-scale structure surrounding the candidate black hole in  $ M87$, the  \cite{EventHorizonTelescope:2019dse}. The study of this compact object via electromagnetic waves found a black hole with a mass of $ M= (6.5\pm0.7)\times 10^{+9} M_{\odot}$ and its spin pointing away from us.
Also, this measurement of the black hole mass in $ M87^{*}$ is consistent with all of the prior mass measurements, and the interpretation of the black hole as a Kerr black hole is not ruled out.
 \\ 
Theoretically, the Kerr-de Sitter metric is used to describe expanding, rotating black holes. This solution was first obtained by Carter \cite{kerr ds2}. It is assumed that rotating black holes are typical objects formed after matter collapses in the universe, and their properties have been extensively studied through observations. The Kerr metric extends the Schwarzschild metric, which was discovered by Karl Schwarzschild in 1915. The Schwarzschild solution describes the geometry of spacetime around an uncharged, spherically symmetric, and non-rotating body. The natural extension to a charged, rotating black hole, the Kerr–Newman metric, was discovered in \cite{Newman:1965tw}. Moreover, the solution for a charged, spherical, non-rotating body, known as the Reissner–Nordström metric, is given in \cite{Reissner, Nordstrom}.\\ 
 Recently, J. Ovalle et al.  interpreted the cosmological
constant $ \Lambda $ as vacuum energy and implemented the so-called gravitational decoupling approach \cite{Ovalle:2021jzf}.A different metric from Carter's has been obtained by \cite{Brahma:2020eos,Abedi:2015yga}, which arises from loop quantum gravity (LQG), a theoretical framework that attempts to reconcile the principles of quantum mechanics with the theory of general relativity. One of the key predictions of LQG is that the area of the event horizon of a rotating black hole is quantized, meaning it can only take on certain discrete values. However, the absence of a rotating black hole model in LQG has been a significant obstacle to testing this prediction through observations. This difficulty limits our ability to compare LQG predictions with observations of real black holes and undermines our capacity to test the theory.

Without a reliable model of rotating black holes in LQG, it is challenging to determine whether the predicted quantization of black hole area is consistent with observations. This hinders the progress of testing LQG against observations and highlights the need for further research to develop a more complete understanding of black holes in the context of LQG.\\
{
Several suggestions have been made in models of loop quantum gravity which may indicate a potential to provide interesting physical effects, such as avoiding singularities encountered in classical general relativity \cite{Bojowald:2015zha}.
Furthermore, it has been noted that quantum space-time effects can change the structure of space-time, and a self-consistent space-time structure should be derived within quantum theories \cite{Bojowald:2018xxu}.
Additionally, recent studies argue that the current models of loop quantum gravity are not fully covariant, indicating the need for non-Riemannian geometry for a consistent description of spacetime \cite{Bojowald:2020unm}. Across these papers, a common issue is the breakdown of general covariance in LQG when quantum corrections are applied.} Furthermore, the general form of conserved quantities related to loop quantum gravity black holes, such as the three laws of thermodynamics, has been derived. In particular, this model fails to comply with the three laws of thermodynamics. Modifying entropy or extending phase space could rescue these conserved quantities \cite{Yang:2023cmv}. \\
The gravitational wave radiation from the motion of the surrounding particles around a polymer black hole in loop quantum gravity may be a potentially useful tool for constraining LQG effects in future gravitational wave detections \cite{Tu:2023xab}.

By using the Newman-Janis algorithm, a polymetric function from the quantum effects of LQG geometry is added to the Kerr geometry. It is shown that the $X$-ray binary GRO J1655-40 establishes an upper bound on it as $0.00086$ at $95\%$ confidence level\cite{Liu:2023vfh}. By  studying the emission from
the accretion flow around Loop Quantum Gravity Black Holes and comparing with the ETH observation of $Sgr A^{\star}$ and $M87^{\star}$, an upper limit for its effective parameter, named as polymetric function $ P $ in LQG,  for $Sgr A^{\star}$
 and  $M87^{\star}$ are found  \cite{Jiang:2023img} while these
blackholes exhibit a preference for a relatively high spin
$ a \geq 0.5$ for $Sgr A^{\star}$ and $0.5 \leq a \leq 0.7$ for $M87^{\star}$.
\\
One way to study gravity in the strong-field regime, such as near black holes, is by using the quasi-periodic oscillations (QPOs). In these phenomena, the radiation from matter falling into a black hole is used as a tool to understand the innermost regions of accretion disks, as well as the masses, radii, and spin periods of black holes. The observed frequencies of QPOs cover a range from millihertz (mHz) up to $0.5$ kilohertz (kHz), and the different types of QPOs are generally divided into low-frequency (LF) QPOs, with a centroid frequency of $30$ Hz, and high-frequency (HF) QPOs, with a centroid frequency of $60$ Hz \cite{Belloni:2009ph}. Typically, QPOs are observed in the X-ray flux emitted by accreting black holes in X-ray binary systems. When a black hole or a neutron star accretes material from a stellar companion, QPOs are observed as narrow features in the power spectra of the light curves from this pair.\cite{Belloni:2009ph, Abramowicz:2001bi,Pasham:2014ybe,alireza}.
The very first hint towards their existence
in the literature dates back to the results reported in \cite{samimi}.

This paper is organized as follows: In Section \ref{II}, we present a simple expression for the geometry under consideration. In Sections \ref{III} and \ref{IV}, we derive the circular timelike orbits in the Brahma, Chen, and Yeom (BCY) background metric, as well as reexamine the ISCO information, including energy, angular momentum, and its radius. In Section \ref{V}, we impose constraints on loop quantum effects using QPO data. In Section \ref{Mw}, we discuss the occurrence of unequal masses for black holes and white holes, providing constraints based on QPOs for this geometry. Finally, Section \ref{sum} is devoted to the summary and discussion.
\section{ The Quantum corrections of the Kerr metric}\label{II}
The theoretically time-reversed counterparts known as "white holes" do not enjoy as much observational support as black holes. The formation of white holes is a intersting topic; one prominent theory suggests that they may arise from black holes, with many of these theories incorporating various quantum mechanical effects\cite{Haggard:2014rza,Bianchi:2018mml,Olmedo:2017lvt,Bodendorfer:2019jay,BenAchour:2020gon,Martin-Dussaud:2019wqc,Rignon-Bret:2021jch,Han:2023wxg,Hong:2022thd,Jalalzadeh:2022rxx}. Recently, a metric was introduced by considering holonomy-corrected effective equations in loop quantum gravity (LQG), utilizing the revised Newman–Janis algorithm. This resulted in a metric that is everywhere non-singular and asymptotically reduces to the Kerr solution\cite{Brahma:2020eos,Abedi:2015yga}. Additionally, the existence of a transition surface induced by non-perturbative quantum corrections can characterize various structures, including a wormhole, a regular black hole featuring an interior spacelike transition surface, or a regular black hole with a timelike transition region located inside the inner horizon.\\

BCY metric takes the following form:
\begin{equation}
    ds^{2}=(1-\dfrac{2M_{0} \beta}{\rho^{2}})dt^{2}-\dfrac{4a M_{0} \beta \sin^{2}{\theta}}{\rho^2} dtd\Phi + \rho^{2} d\theta^{2}+ \dfrac{\rho^2}{\Delta} dr^2 + \dfrac{\Sigma \sin^2 \theta}{\rho^2} d\Phi^2
\end{equation}
where 
\begin{eqnarray}
    \Delta &=& 8 \lambda M_{B}^{2} A \beta^{2}+ a^2  \nonumber\\
    \Sigma&=&(\beta^{2}+a^{2})-a^{2} \Delta\ \sin^{2}\theta \nonumber\\
    M_{0}&=&\dfrac{1}{2} \beta(1-8\lambda m_{B}^{2} A)\\ \rho^{2}&=&\beta^2 + a^2 \cos^{2} \theta \nonumber
\end{eqnarray}
with 
\begin{eqnarray}
    \beta^2&=&\dfrac{\lambda}{\sqrt{1+x^2}}\dfrac{M_{B}^2 (x+\sqrt{1+x^2})^{6}+M_{W}^{2}}{(x+\sqrt{1+x^2})^{3}} \nonumber\\
    A&=&(1-\dfrac{1}{\sqrt{2\lambda}}\dfrac{1}{\sqrt{1+x^2}}) \dfrac{1+x^{2}}{\beta^{2}} \nonumber\\
\end{eqnarray}
and $ x=\dfrac{r}{M_{B}\sqrt{8\lambda}}$, $M_B$ and $M_W$ 
 are the mass of asymptotically Kerr blackhole and white hole, respectively. Also $\lambda= \dfrac{[\lambda_{\kappa}/(M_{B} M_{W})]^{2/3}}{2}$ is the dimensionless and non-negative parameter, $ \lambda_\kappa$ is a quantum parameter originated from holonomy modifications\cite{Bodendorfer:2019nvy}.

 \subsection{Equal mass black hole and white hole }
 Here we show that when $M_{B}=M_{W}=M$ the metric takes simpler form. We start with the expression for $\beta$
 $$\beta^2=\dfrac{\lambda}{\sqrt{1+x^2}}\dfrac{M_{B}^2 (x+\sqrt{1+x^2})^{6}+M_{W}^{2}}{(x+\sqrt{1+x^2})^{3}}$$
Let's define $y=x+\sqrt{1+x^2}$ and set $\kappa=\frac{M_{w}}{ M_{B}}$. We get:
\begin{equation}
\beta^2=\dfrac{2M^2\lambda}{1+y^2}\dfrac{ y^{6}+\kappa^{2}}{y^{2}}
\end{equation}
Now, by using the identity:
\begin{equation}
1+y^6=\left(y^2+1\right) \left(y^4-y^2+1\right)
\end{equation}
We can rewrite $\beta$ as:
\begin{equation}
\beta^2=\dfrac{2M^2\lambda}{y^2} {( y^4-y^2+1)}+\dfrac{2M^2\lambda}{1+y^2}\dfrac{ \kappa^{2}-1}{y^{2}}
\end{equation}
or equivalently in terms of $x$ as:
\begin{equation}
\beta^2={2M^2\lambda} {(1+4 x^2)}+{2M^2\lambda}{(\kappa^{2}-1)} \frac{ \left(1-\frac{{x}}{\sqrt{{x}^2+1}}\right)}{\left(\sqrt{{x}^2+1}+{x}\right)^2}
\end{equation}
when $M_w=M_B$ or $\kappa=1$ we get a simple form for the metric. We have
$$\beta^2={2M^2\lambda} {(1+4 x^2)}$$
Therefore the metric becomes:
 \begin{equation}\label{newmetric}
    ds^{2}=(1-\dfrac{2M_{0}r}{\rho^{2}})dt^{2}-\dfrac{4a M_{0} r\sin^{2}{\theta}}{\rho^2} dtd\Phi + \rho^{2} d\theta^{2}+ \dfrac{\rho^2}{\Delta} dr^2 + \dfrac{\Sigma \sin^2 \theta}{\rho^2} d\Phi^2
\end{equation}
where 
 \begin{eqnarray}
    \Delta &=& r^2+ a^2 -2 M_0 r +2 \lambda M^2 \nonumber\\
    \Sigma&=&(r^2+2 \lambda  {M}^2+a^{2})^2-a^{2} \Delta\ \sin^{2}\theta \nonumber\\
    M_{0}&=&\tfrac{M}{r} \left(\sqrt{8 \lambda  M^2+r^2}-3 \lambda  M\right)\\ \rho^{2}&=&r^2+ a^2 \cos^{2} \theta+2 \lambda  {M}^2  \nonumber
\end{eqnarray}
The metric presented in Eq.~\eqref{newmetric} is more conducive to computation and comparison with the Kerr metric. However, this simplified form is only applicable in situations where $\kappa=1$ , or $M_w= M_B $. In the following sections, we will examine this case, and in Section \ref{Mw}, we will explore the effects of the inequality $M_w\neq M_B$. 
\section{ISCO for BCY Black hole }\label{III}
he study of ISCOs (Innermost Stable Circular Orbits) is crucial because they represent the final stable orbits for matter around a black hole. Research has focused on the behavior of ISCOs for both non-spinning and spinning particles across various black hole backgrounds, including Schwarzschild, Kerr, and Kerr-Newman (KN). Numerous studies have investigated these phenomena, such as: \cite{Bardeen:1972fi,Carroll:1997ar,Suzuki:1997by,Zhang:2017nhl,Jefremov:2015gza}.\\
The metric coefficients are still independent of $t$ and $\phi$ coordinates. Thus $t$ and $\phi$ components of the 4-velocity of test particles can be written as: \cite{Bambi:2013fea}
\begin{eqnarray}
\dot{t}=\dfrac{g_{\phi\phi} \mathcal{E}+g_{t\phi} \mathcal{L}}{g_{t\phi}g_{\phi t}-g_{tt}g_{\phi\phi}}~~~~~,~~~~~\dot{\phi}=\dfrac{g_{t\phi} \mathcal{E}+g_{tt} \mathcal{L}}{g_{tt}g_{\phi\phi}-g_{t\phi}g_{\phi t}}
\end{eqnarray}
where $\mathcal{E}$ and $\mathcal{L}$ are conserved quantities related to energy and  $z$ component of angular momentum at infinity respectively. From the conservation of the rest-mass, $ g_{\mu\nu}\dot{x}^{\mu} \dot{x}^{\nu}=-1$, we have: $g_{rr}\dot{r}^{2}+g_{\theta\theta}\dot{\theta}^{2}= V_{eff}(r,\theta,\mathcal{E},\mathcal{L})$. Therefore, for this metric, the effective potential of the orbits is given by:
\begin{equation}\label{potential}
\dot{r}^{2}=V_{eff}(r,\theta,\mathcal{E},\mathcal{L})=\dfrac{\mathcal{E}^{2}g_{\phi\phi}+2\mathcal{E}\mathcal{L}g_{t\phi}+\mathcal{L}^{2}g_{tt}}{g_{t\phi}^{2}-g_{tt}g_{\phi\phi}}-1
\end{equation}

The innermost stable circular orbit (ISCO) is the smallest marginally stable circular orbit in which a test particle can orbit a massive object. The ISCO plays a crucial role in black hole accretion disks, as it defines the inner edge of the disk. The radius of the black hole’s ISCO can be determined in the equatorial plane using
constant r, i.e.  $ \dot{r}^{2} = 0$ and $\dfrac{d V_{eff}(r,\theta,\mathcal{E},\mathcal{L})}{dr} = 0$   and fixing $\theta=\tfrac{\pi}{2}$. The radial component of equations for ISCO   is given by:
\begin{align}
\dot{r}^2=&-1+\mathcal{E}^2+\frac{a^2 \left(\mathcal{E}^2-1\right)+2 M \left(\sqrt{8 \lambda  M^2+r^2}-3 \lambda  M\right)-\mathcal{L}^2}{2 \lambda  M^2+r^2}\\&+\frac{2 M (\mathcal{L}-a \mathcal{E})^2 \left(\sqrt{8 \lambda  M^2+r^2}-3 \lambda  M\right)}{\left(2 \lambda  M^2+r^2\right)^2}\nonumber
\end{align}
With the condition $ \dfrac{d V_{eff}}{dr}=0 $, the specific energy, $ \mathcal{E} $ and the angular momentum $\mathcal{L} $ of the particle in the ISCO around the new blackhole to the  first order in $a$ and  $\lambda$ are 
\begin{figure}
    \centering
\includegraphics[scale=.7]{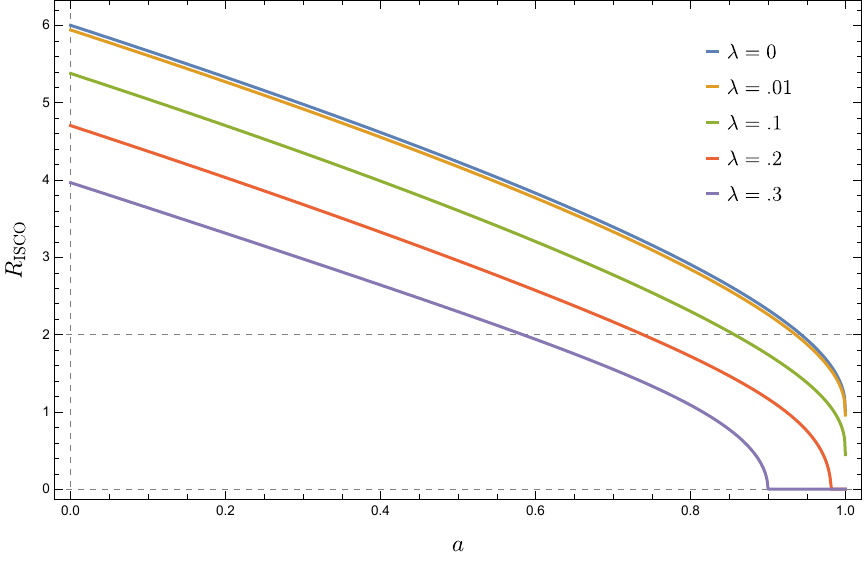}
    \caption{The effect of $\lambda$ in $R_{ISCO}$ in terms of $a$. Scales are such that $M=1$. As $\lambda$ increases, the radius of ISCO decreases. }
    \label{fig:enter-label}
\end{figure}
\begin{align}
   & \mathcal{E}= \frac{2 \sqrt{2}}{3}(1-\frac{\lambda}{18})-\frac{a}{18 \sqrt{3} M} (1+\frac{71\lambda}{36})\\
   & \mathcal{L}= 2 \sqrt{3} M (1-\frac{23\lambda}{36})-\frac{2 \sqrt{2}}{3}  a (1+\frac{5\lambda}{6})
\end{align}
{It is possible to find an analytical expression for the radius of ISCO in the case of small values of $\lambda$ and $a$ (slow-rotating black holes). The result   to first order in $a$ and $\lambda$ is}
\begin{eqnarray}
 \dfrac{r}{M} =6(1-\lambda)-4\dfrac{a}{M}\sqrt{\frac{2}{3}}(1+\frac{\lambda}{12})
\end{eqnarray}
This equation will reduce to the Schwarzschild result when $\lambda=a=0$.
In the Schwarzschild
background,  as a non-rotating black hole, the radius of ISCO equals to $6M$ \cite{Kaplan,Landau}.  { On the other hand it agrees with the result of \cite{Tu:2023xab} for non-rotating black holes in quantum gravity.} However, for rotating black holes, the values of ISCO parameters—such as radius, total angular momentum, energy, and orbital angular frequency—depend on the direction of particle motion in relation to the direction of black hole rotation. In the case
of the extreme Kerr background, we have $9M$ for
the anti-parallel and $M$ for the parallel orientation vectors of the orbital angular momentum of the particle \cite{Landau,Ruffini,Bardeen:1972fi}.  We find that the ISCO radius decreases as we increase $\lambda$ as shown in Fig.~\ref{fig:enter-label}. This result is consistent with the previous work \cite{Tu:2023xab}.
\section{QPOs and Resonance  conditions}\label{IV}
Let us now examine nearly circular and equatorial orbits for non-spinning particles with small oscillation frequencies. To facilitate this analysis, the circular orbit of radius
$R$ in the equatorial plane can be described by a straightforward set of conditions for the 4-velocity vector, as follows: $\dot{r}=\dot{\theta}=0$. \\
 The geodesic equation  gives:
\begin{equation}\label{geodesic}
\Omega_\phi=\frac{\dot{\phi}}{\dot{t}}=\frac{-\partial_r g_{t \phi} \pm \sqrt{\left(\partial_r g_{t \phi}\right)^2-\left(\partial_r g_{t t}\right)\left(\partial_r g_{\phi \phi}\right)}}{\partial_r g_{\phi \phi}}
\end{equation}
Since the metric coefficients are independent of the $t$ and $\phi$ coordinates, we have the conserved specific energy at infinity, $E$ and the conserved angular momentum at infinity, $J$. They are given by 
\begin{eqnarray}
E(\Omega_{\phi})&=& -\dfrac{g_{tt}+ g_{t\phi}\Omega_{\phi}}{\sqrt{-g_{tt}-2g_{t\phi}\Omega_{\phi}-g_{\phi\phi}\Omega_{\phi}^2}}\\&=&\frac{a+(r-2) \sqrt{r}}{\sqrt{2 a r^{3/2}+(r-3) r^2}}+\frac{\lambda  \left(-3 a^2 (r-1)+a \sqrt{r} (9 r-5)-4 r^2\right)}{r^2 \left(2 a+(r-3) \sqrt{r}\right) \sqrt{2 a r^{3/2}+(r-3) r^2}}+O\left(\lambda ^2\right)\nonumber
\end{eqnarray}
and
\begin{eqnarray}
J(\Omega_\phi)&=&  \dfrac{g_{t\phi}+ g_{\phi\phi}\Omega_{\phi}}{\sqrt{-g_{tt}-2g_{t\phi}\Omega_{\phi}-g_{\phi\phi}\Omega_{\phi}^2}}\nonumber\\&=&\frac{a^2-2 a \sqrt{r}+r^2}{\sqrt{2 a r^{3/2}+(r-3) r^2}}
\\&+&\frac{\lambda\left(-3 a^3 (r-1)+a^2 \sqrt{r} (-3 (r-4) r-5)+3 a (r-1) r^2+r^{5/2} ((8-3 r) r-9)\right)}{r^2 \left(2 a+(r-3) \sqrt{r}\right) \sqrt{2 a r^{3/2}+(r-3) r^2}}+O\left(\lambda ^2\right)  \nonumber
\end{eqnarray}
To find the radial and vertical epicyclic frequencies, we need to calculate the small perturbations around circular orbits. We write the perturbations as $r(t)=R+\delta r$ and $\theta(t)=\dfrac{\pi}{2}+\delta \theta$ where the equations governed by  $\delta r$, $\delta \theta$ are  given by \cite{Colistete:2002ka}
 \begin{equation}\label{eq: osicllation r}	
		\frac{d^2 \delta r}{ds^2} + \Omega_{r}^{2}\delta r=0 
	\end{equation}
 \begin{equation}\label{eq: osicllation theta}
		\frac{d^2 \delta \theta }{ds^2}+ \Omega_{\theta}^{2} \delta\theta=0 
	\end{equation}
	where
	\begin{eqnarray}\label{28}
	\Omega_{r}^{2}=\dfrac{1}{2 g_{rr}\dot{t}^2}\dfrac{\partial^{2}V_{eff}}{\partial r^{2}}\vert_{\theta=\dfrac{\pi}{2}}~~~~~,~~~~~\Omega_{\theta}^{2}=\dfrac{1}{2 g_{\theta\theta}\dot{t}^2}\dfrac{\partial^{2}V_{eff}}{\partial \theta^{2}}\vert_{\theta=\dfrac{\pi}{2}}
	\end{eqnarray}
Using the expression for $V_{eff}$, e.g. Eq.(\ref{potential}) and
\begin{align}
\dot{t}=\frac{1}{\sqrt{-g_{t t}-2 g_{t \phi} \Omega_\phi-g_{\phi \phi} \Omega_\phi^2}}  
\end{align}
 these  $\Omega_r$ and $\Omega_{\theta}$ can be derived. Due to the length of these terms, we will not include them here.  We present the terms linear in $\lambda$, therefore the frequencies for orbits take the following form

\begin{eqnarray}\label{1}
  \frac{\Omega_r^2}{\Omega_\phi^2}  &=&\frac{-3 a^2+8 a \sqrt{M} \sqrt{r_{c}}+r_{c} (r_{c}-6 M)}{r^2}\\&+&\frac{2 \lambda  M \left(-3 a^2 (M-r_{c})+4 a \sqrt{M} \sqrt{r_{c}} (M-3 r_{c})+r_{c} \left(6 M^2-M r_{c}+3 r_{c}^2\right)\right)}{r_{c}^4}+O\left(\lambda ^2\right)\nonumber
\end{eqnarray}
and
\begin{align}\label{2}
  \frac{\Omega_\theta^2}{\Omega_\phi^2}  &=\left(\frac{3 a^2}{r_{c}^2}-\frac{4 a \sqrt{M}}{r_{c}^{3/2}}+1\right)+\frac{6 a \lambda  M \left(a (r_{c}-3 M)+2 M^{3/2} \sqrt{r_{c}}\right)}{r_{c}^4}+O\left(\lambda ^2\right)
\end{align}
  \begin{figure}[ht]
\centering
\subfloat[]{\includegraphics[width=7.4cm]{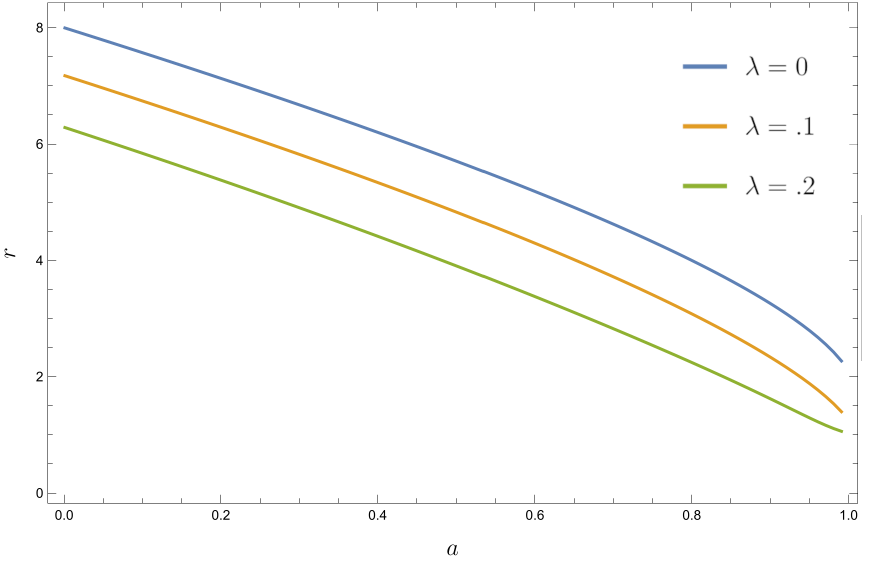}}
\qquad
\subfloat[]{\includegraphics[width=7.4cm]{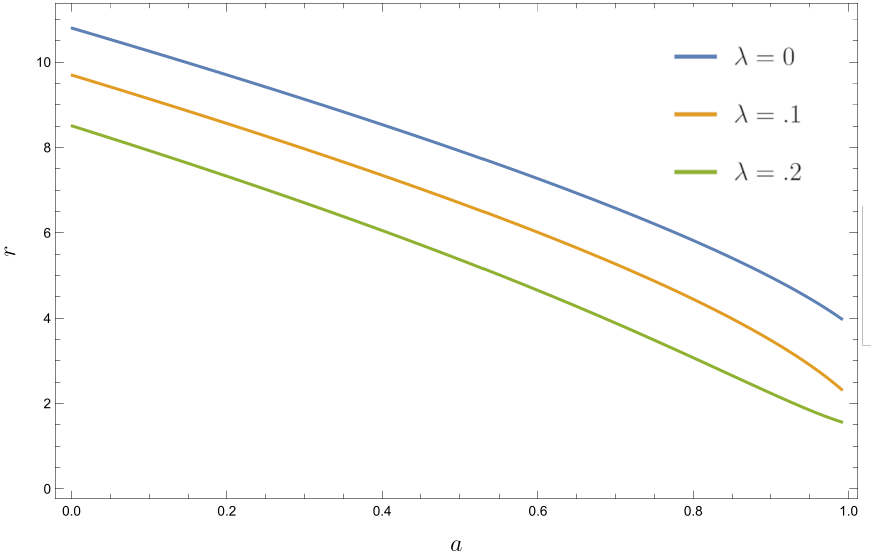}}
\qquad
\subfloat[]{\includegraphics[width=7.4cm]{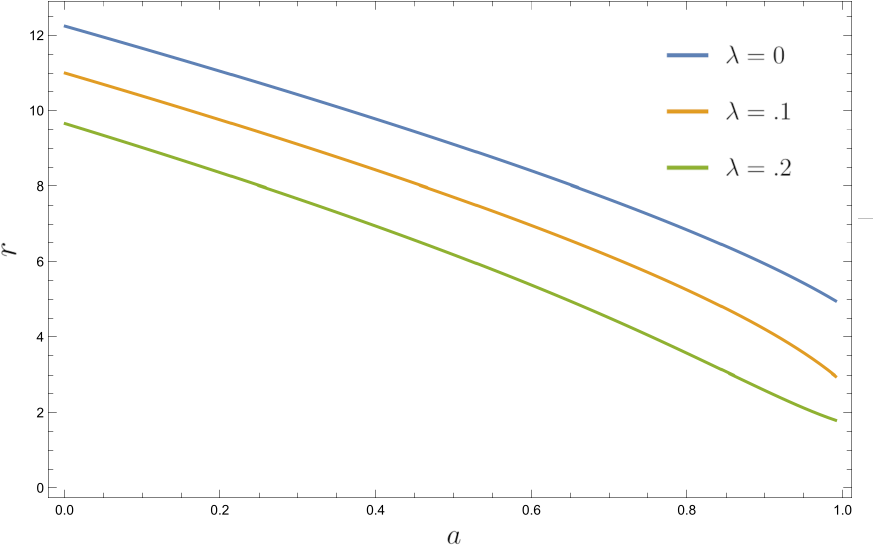}}
\caption{ The location of  $r\theta$ resonant orbits  for different values of $\lambda$ parameter and blackhole's momentum. The
$(a)$, $(b)$and $(c)$ are related to $1 : 2$, $2 : 3$ and $5 : 7$ resonance order respectively.  We have used Eq. (\ref{1}), and Eq. (\ref{2}) to obtain the figures.
}
\label{rtheta}
\end{figure}
 \begin{figure}[ht]
\centering
\subfloat[]{\includegraphics[width=7.8cm]{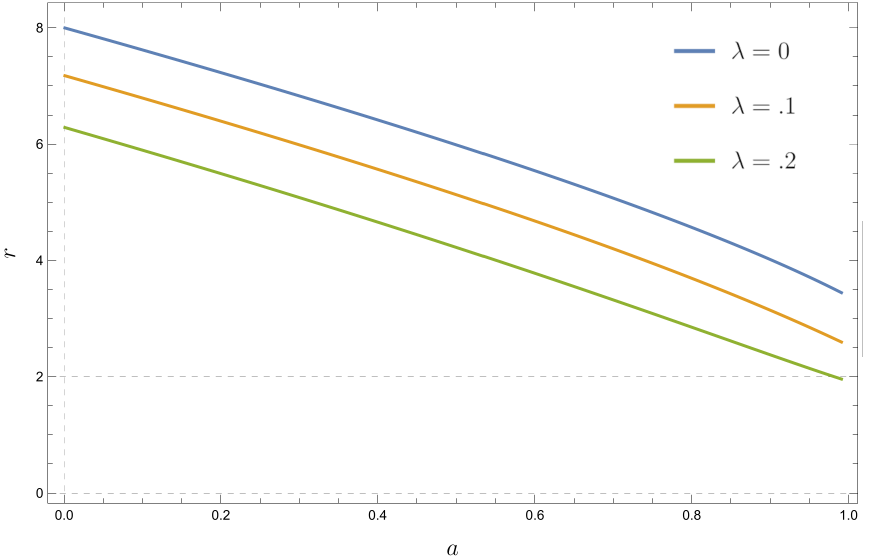}}
\qquad
\subfloat[]{\includegraphics[width=7.8cm]{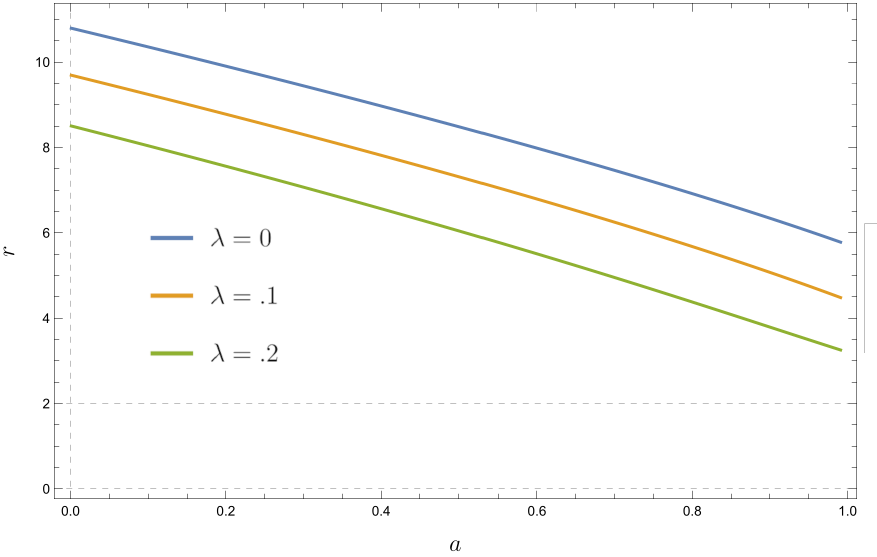}}
\qquad
\subfloat[]{\includegraphics[width=7.8cm]{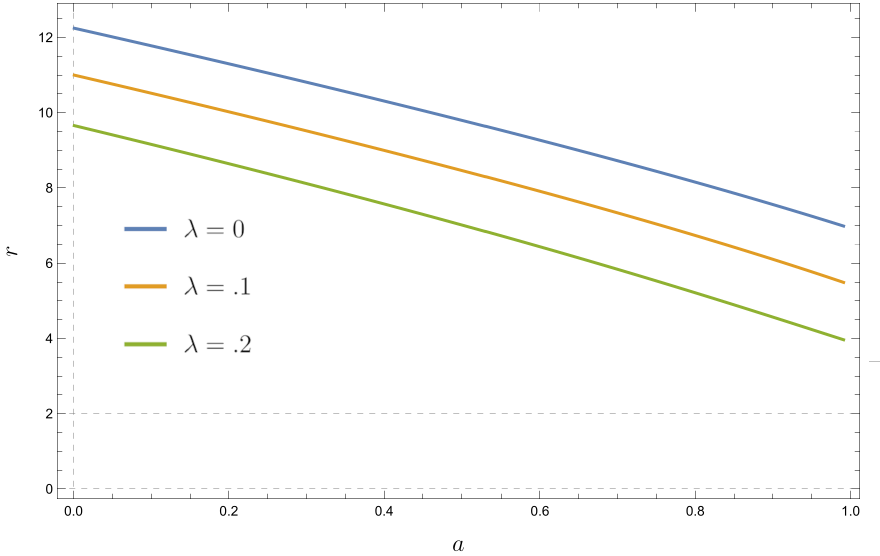}}
\caption{ The location of  $r\phi$ resonant orbits  for different values of $\lambda$ parameter and black hole's momentum. The
$(a)$, $(b)$and $(c)$ are related to $1 : 2$, $2 : 3$ and $5 : 7$ resonance order respectively. Here, we have used Eq. (\ref{1}), and Eq. (\ref{2}) to obtain the 
figures.
}
\label{rphi}
\end{figure}
 \begin{figure}[ht]
\centering
\subfloat[]{\includegraphics[width=7.4cm]{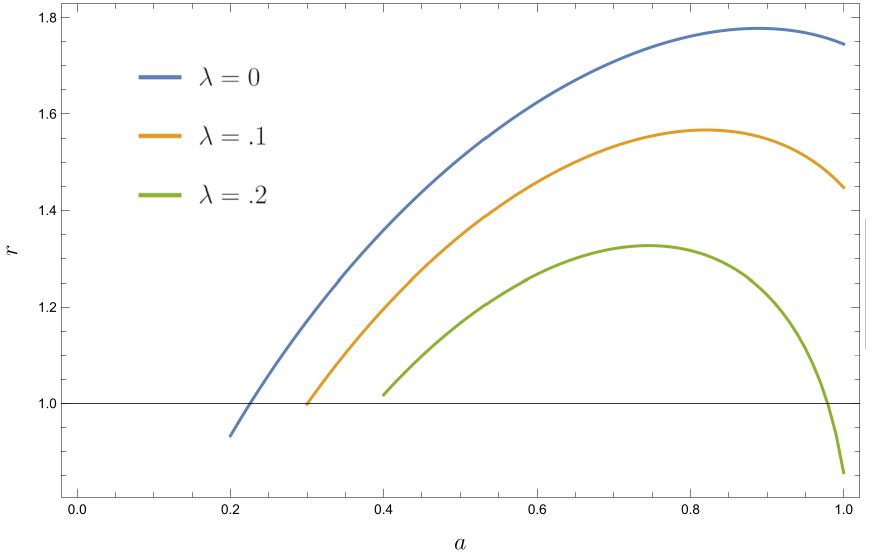}}
\qquad
\subfloat[]{\includegraphics[width=7.4cm]{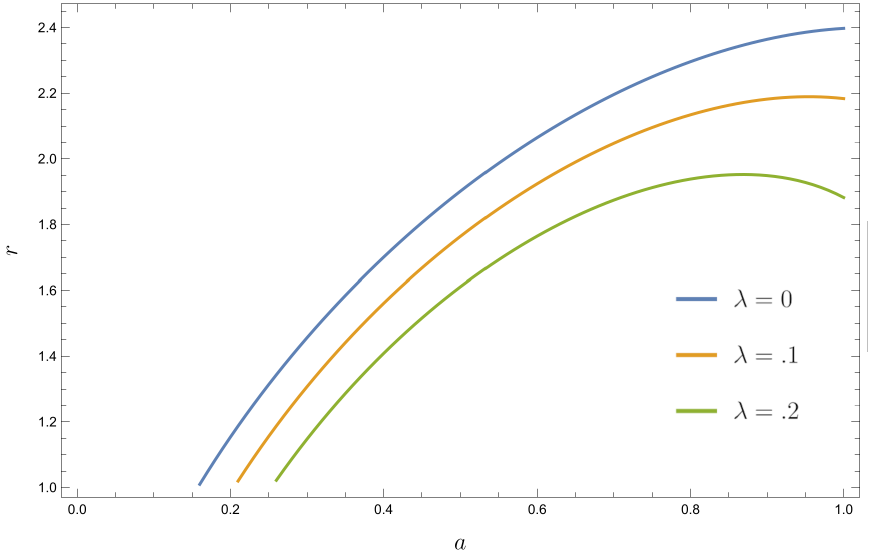}}
\qquad
\subfloat[]{\includegraphics[width=7.4cm]{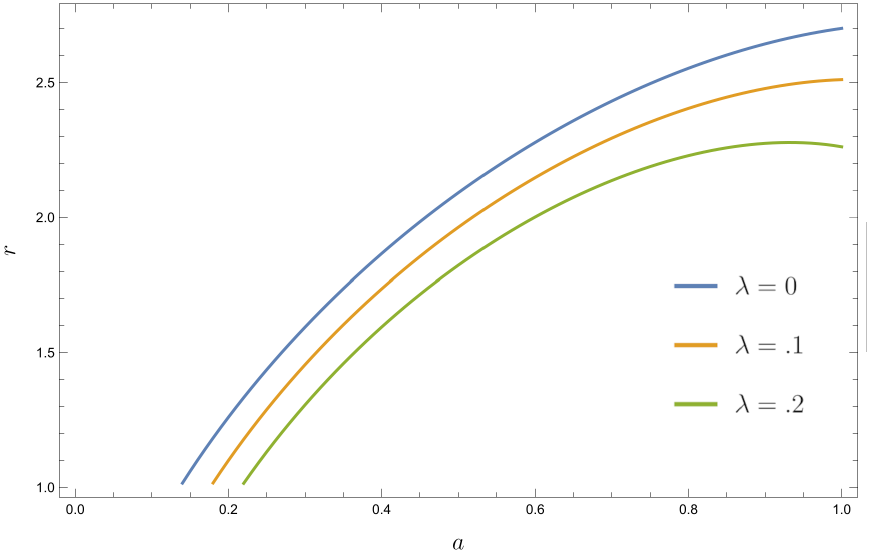}}
\caption{ The location of  $\theta\phi$ resonant orbits   for different values of $\lambda$ parameter and black hole's momentum. The
$(a)$, $(b)$and $(c)$ are related to $1 : 2$, $2 : 3$ and $5 : 7$ resonance order respectively. 
}
\label{thetaphi}
\end{figure}
 \begin{figure}
    \centering
\includegraphics[scale=.8]{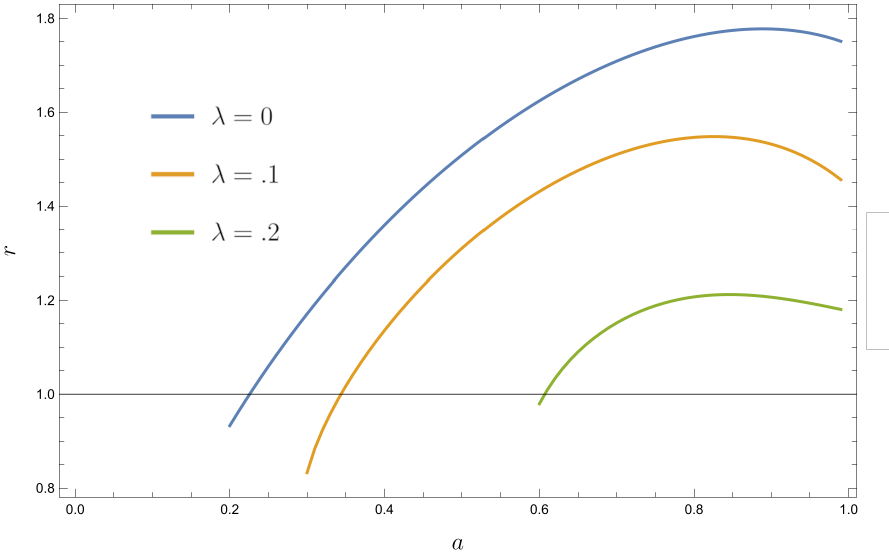}
MNM    \caption{The location of $\theta\phi$ resonant orbits for different values of $\lambda$ parameter. }
    \label{fig:ra}
\end{figure}
is the radius of the circular orbit of a particle undergoing small oscillations. The general condition for resonance can be expressed in terms of the fundamental frequencies as $ \alpha \Omega_{r}  + \beta \Omega_{\theta} + \gamma \Omega_{\phi} =0$ where $ \alpha $, $ \beta $ and $ \gamma $ are rational numbers. For circular orbits, $  \Omega_{r} $ is not particularly relevant or in limiting case on the equatorial plane $ \Omega_{\theta} $ has no meaning, while the other frequencies play a key role in the orbital dynamics.\\  We can now demonstrate the resonance orbits depending on the black hole's angular momentum and LQG parameters. In Figs. \ref{rtheta}, \ref{rphi}, and \ref{thetaphi} resonance orbits are depicted as a function of black hole's angular momentum, while the LQG parameter takes different values. By increasing the LQG parameter the orbits are getting closer to the event horizon and this is true for each of the resonance frequencies. Furthermore, both $r\theta$ and $r\phi$ exhibit identical behavior, while $\theta\phi$ behaves differently. In the case of $\theta\phi$ resonance, the orbit initially moves away from the horizon, but as we increase the angular momentum of the black hole, it gets closer to the horizon again. 
\section{Constraints by QPOs}\label{V}
The radial and vertical epicyclic frequencies are defined as $ \nu_{r}=\dfrac{\Omega_{r}}{2\pi}$ and $ 
\nu_{\theta}=\dfrac{\Omega_{\theta}}{2\pi}$ and the periastron precession frequency $ \nu_{p} $ and the nodal precession frequency $ \nu_{n} $  are also defined as 
\begin{equation}
\nu_{p}=\nu_{\phi}-\nu_{r}~~~~~,~~~~~\nu_{n}=\nu_{\phi}-\nu_{\theta}
\end{equation} 
 Typically these frequencies depend on the Kerr parameters, the blackhole mass, $M$ its spin $a$, and the radius of the orbit $r$. When new parameters are added to the metric, these frequencies will also depend on these parameters. The orbital frequency $ \nu_{\phi} $ corresponds to the observed upper high-frequency of QPOs, the periastron precession frequency $ \nu_{p} $  corresponds to the observed lower high-frequency QPOs  and the nodal  precession frequency $ \nu_{n} $ corresponds to the observed lower-frequency  QPOs \cite{Motta:2013wga,Belloni:2016xgi}.\\
 
Our aim is to fit the metric proposed in the previous section to set limits for $\lambda$  using the data from GRO~J1655$-$40. GRO~J1655$-$40 is an X-ray binary system where one of the stars could be a black hole.The X-ray timing method has provided three QPO frequencies by RXTE observations. We use a Bayesian approach to constrain the parameters. We assume that the likelihood is given by $\cal
L\sim\rm e^{-\chi^2/2}$, where the $\chi$-square takes the following form
\begin{eqnarray}
\chi^{2}(a,\lambda,M,r)=& 
\frac{\left( \nu_{\rm C} - \nu_{\rm n} \right)^2}{\sigma^2_{\rm C}}
+ \frac{\left( \nu_{\rm L} - \nu_{\rm p} \right)^2}{\sigma^2_{\rm L}}
+ \frac{\left( \nu_{\rm U} - \nu_\phi \right)^2}{\sigma^2_{\rm U}} \, ,
\label{eq-chi20}
\end{eqnarray} 
where $\nu_{\rm C}$, $\nu_{\rm L}$ and $\nu_{\rm U}$, as well as their errors given
by $\sigma^2_{i}$ ($i \in \{\rm C, L, U\}$), are provided by the
observations.
For GRO~J1655$-$40 these frequencies are measured and we have \cite{Motta:2013wga}
\begin{align}
\begin{matrix}
\nu_{\rm C} = 17.3 \; {\rm Hz} \, , && \sigma_{C} = 0.1 \; {\rm Hz} \, , \\
\nu_{\rm L} = 298 \; {\rm Hz} \, , && \sigma_{L} = 4 \; {\rm Hz} \, , \\
\nu_{\rm U} = 441 \; {\rm Hz} \, , && \sigma_{U} = 2 \; {\rm Hz} \, . \\
\end{matrix}
\label{tab1}
\end{align}
We define the
posterior as
\begin{align}
\mathcal{P}\left(a/M,M, \lambda,r|\nu_i \right)\propto \mathcal{L}\, p(\lambda) \, p(a/M) \, p(M) \,p(r),
\end{align}
We use the mass measurement as our  prior and set $p(M/M_\odot)\sim\rm e^{-\frac{1}{2} \left(\frac{M/M_\odot-5.4}{0.3}\right) ^{2}}$ throughout this work.
The covariance between variables is illustrated in Fig.~\ref{fig:qpo}. We find that $\lambda=0.15^{+0.23}_{-0.14}$.
We see that there is degeneracy between  $\lambda$ and M in the measurement and it is a combination of these parameters which can be measured with more precision. $\lambda$ and $M$ are also correlated. Our result does not exclude the Kerr metric as the possible candidate because the Kerr metric lies in the $68\%$ credible interval. {In terms of $\lambda_k$, the quantum gravity scale parameter, we see that in Fig.~\ref{fig:qpo} the most probable value agrees with the Kerr blackhole. If we use the average of posterior    $\lambda=0.15$ and $M=6.28 M_{\odot}$, we estimate that $\sqrt{\lambda_k}< 3.7\, \texttt{Km}$. The Schwarzschild radius for this blackhole is almost $16 \, \texttt{Km} $. The quantum gravity scale is smaller than the Schwarzschild radius. This bound is weaker than the bound reported in \cite{Ashtekar:2021kfp}.}
\begin{figure}[h!]
    \centering
\includegraphics[scale=0.7]{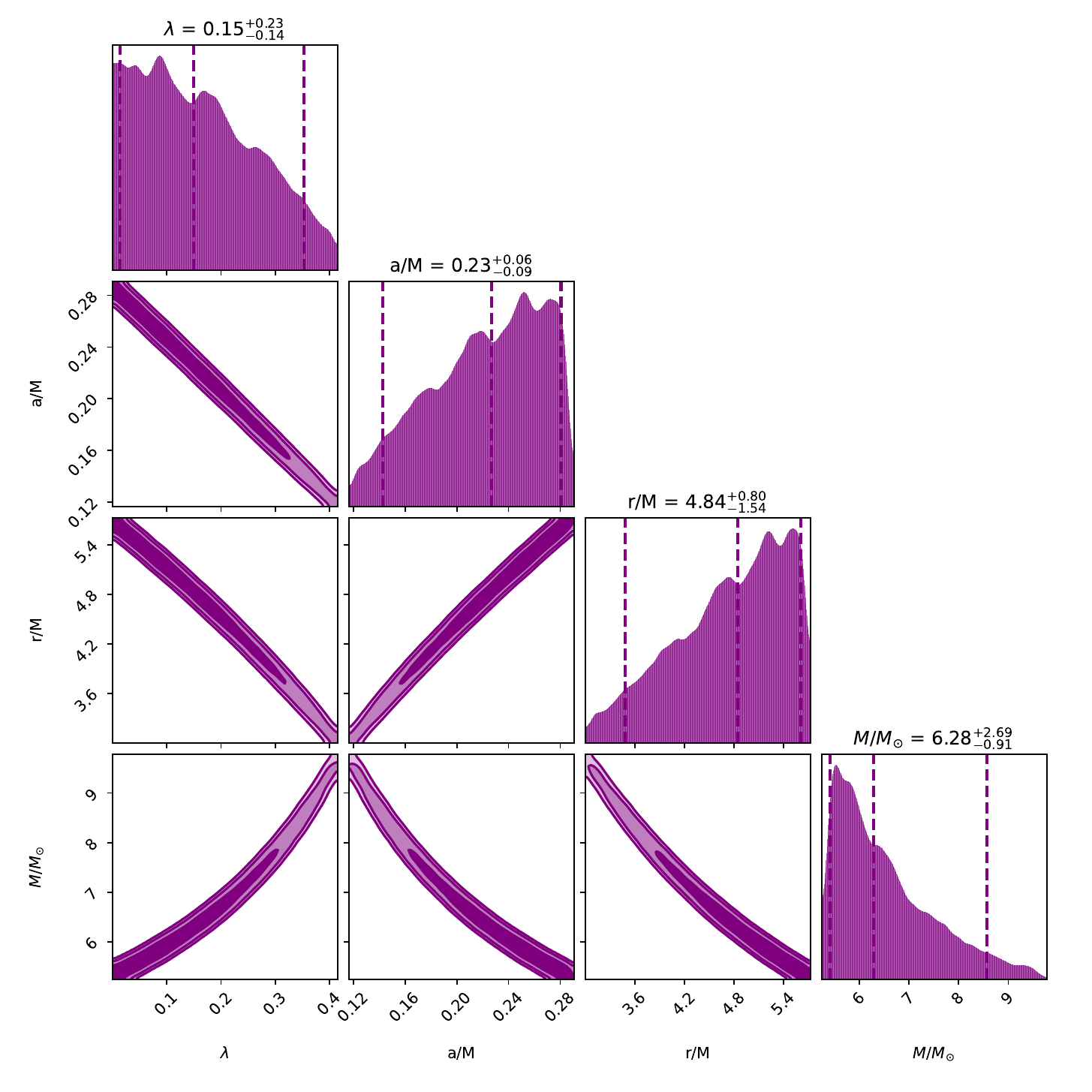}  
\caption{Nonlinear QPOs, the contours show 68\%, 95\% and
99\% credible intervals. Dashed lines are the 5\%, 50\%, and 95\% percentiles of the distribution. We see there is a degeneracy between $\lambda$,$M$, $r$ and $a$. It is a combination of $\lambda$ and $M$ which can be constrained with more precision. }
    \label{fig:qpo}
\end{figure}
\begin{figure}[H] 
\includegraphics[scale=0.7]{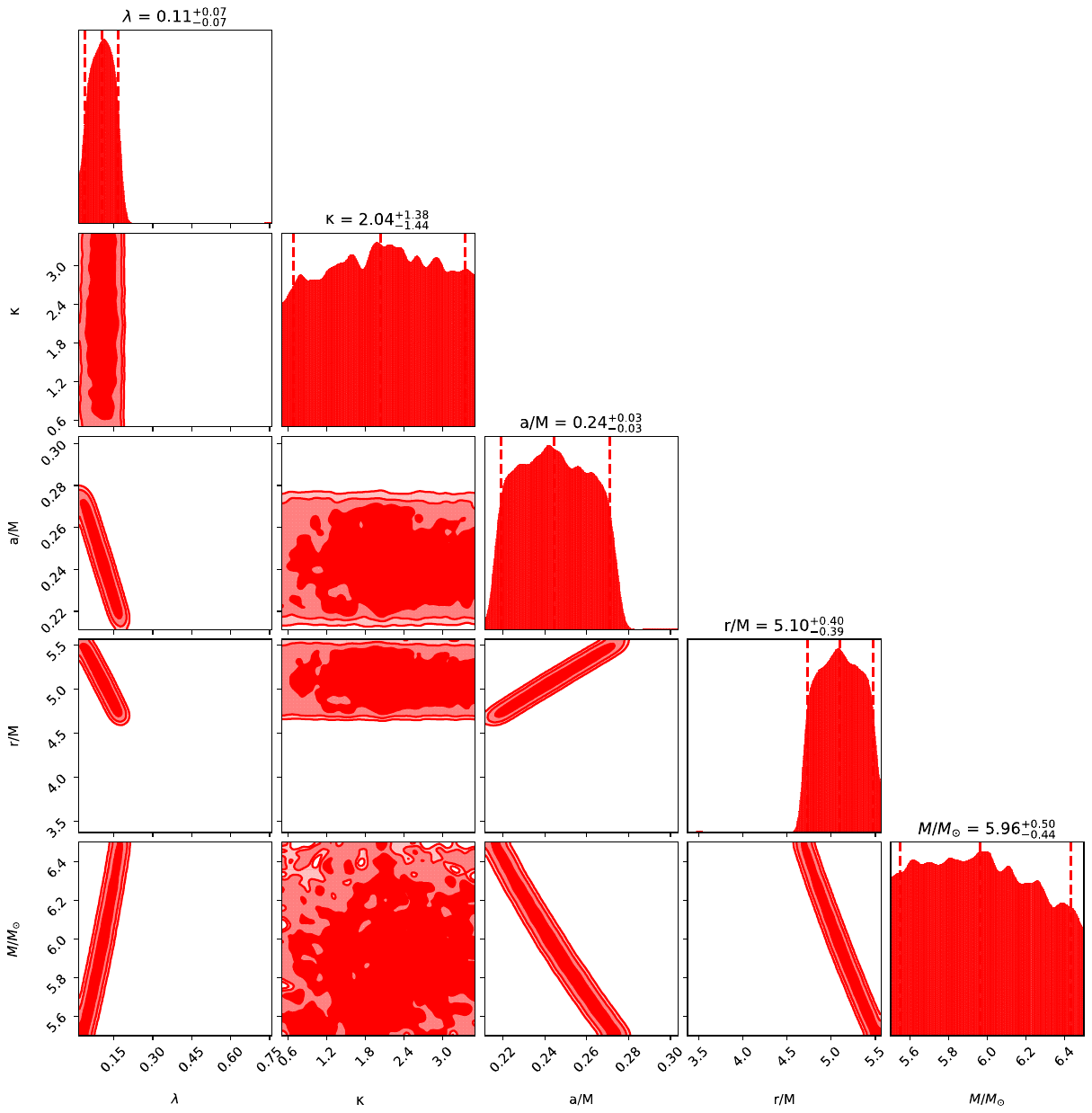}  
\caption{Nonlinear QPOs for the case of unequal mass blackhole white hole, $\kappa \neq 1$. The contours show 68\%, 95\% and
99\% credible intervals. Dashed lines are the 5\%, 50\%, and 95\% percentiles of the distribution. We see that  $\lambda$, $M$, $r$ and $a$ are degenerate. It is a combination of $\lambda$ and $M$ which can be constrained with more precision.}
    \label{fig:qpo1}
\end{figure}

\section{Unequal mass blackhole and white hole: the effect of $M_w$}\label{Mw}
In this section, we consider the case where the mass $M_w$ and $M_B$ are not necessarily equal or $\kappa\neq 1$. The approach will be the same as in the previous section. The general expression for $\Omega_\phi$ is very complicated. However, as the value of $\lambda$ is expected to be small $\lambda\ll1$, we can expand the expressions to the lowest orders in $\lambda$.  Then the $\Omega_\phi$, $\Omega_r$ and $\Omega_{\theta}$  will be given as:
\begin{align}
    \Omega_\phi&= \Omega_\phi^{(0)}+\Omega_\phi^{(1)}\lambda+\Omega_\phi^{(2)}\lambda^2+\Omega_\phi^{(3)}\lambda^3+\mathcal{O}(\lambda^4)\nonumber\\
      \Omega_r&= \Omega_r^{(0)}+\Omega_r^{(1)}\lambda+\Omega_r^{(2)}\lambda^2+\Omega_r^{(3)}\lambda^3+\mathcal{O}(\lambda^4)\nonumber\\
        \Omega_\theta&= \Omega_\theta^{(0)}+\Omega_\theta^{(1)}\lambda+\Omega_\theta^{(2)}\lambda^2+\Omega_\theta^{(3)}\lambda^3+\mathcal{O}(\lambda^4)\nonumber\\
\end{align}
The $\Omega_\phi^{(i)}$  $\Omega_r^{(i)}$, and  $\Omega_\theta^{(i)}$ are given in the appendix \ref{omegas}. 
One important aspect of the expressions is that the effect of $M_w$ just appears in the third order of expansion in $\lambda$ parameter.

\subsection{Constraints by QPOs}

We use the QPO observation to constrain the model. Because the model has one more parameter and we have three QPO's measurement, we expect more degeneracy between parameters. The results are presented in Fig.~\ref{fig:qpo1}. It turns out that $\kappa$ is not constraint by QPO experiments. We find an improvement on the constraint of $\lambda$, where we have $\lambda=0.11^{+0.07}_{-0.07}$. This model hints that  the compact object  may not be Kerr metric. However, the other parameter is not constrained with much precision  $\kappa=2.04^{1.38}_{1.44}$. This result also includes the case where $\kappa=1$. Thus, the equal mass blackhole white hole is allowed at $1\sigma$ credible interval. {We again use the values in Fig.~\ref{fig:qpo1} to write the result for   $\lambda_k$. We find that $\sqrt{\lambda_k}< 3.7 \, \texttt{Km}$. This results is still weaker than the bound in \cite{Ashtekar:2021kfp}.}

\section{Summary and discussion}\label{sum}
We study two geometries arising from quantum gravity corrections that describe a rotating black hole. In the first scenario, the black hole and white hole can possess equal mass. In the second scenario, we consider black holes and white holes with unequal masses.
 We provide a simpler definition of the presence of the LQG parameter in the Kerr metric. We derived ISCO's in terms of, $\lambda $ using its effective potential and found that by increasing $\lambda$, the radius of ISCO decreases. We obtain two relative fundamental frequencies for a rotating black hole then by using $X-$ray binary $GRO J1655-40$ (Fig.\ref{fig:qpo}) we find the bound on the polymetric function as $\lambda=0.15^{+0.23}_{-0.14}$ at $68\%$ confident level. In the case of a black hole and white hole with unequal masses, we find that $ \kappa=2.04^{+1.38}_{-1.44} $ and  $\lambda=0.11^{+1.38}_{-1.44}$ at $1\sigma$ confidence level. {In terms of $\lambda_k$, we get the upper bound $\sqrt{\lambda_k}< 3.7 \, Km$.  \\
}
We find that $M$  and $\lambda$ are degenerate  highly correlated, which hinders our ability to constrain $\lambda$  with precision.  highly correlated, which hinders our ability to constrain

We employed test mass orbits within the relativistic precession model. However, complications arise in modeling the accreting material, which could impact the quasi-periodic oscillations (QPOs). Additionally, environmental factors, such as thermal effects and the optical thickness of the accreting material or gas emissions, may also influence the behavior of the QPOs.

\appendix
\section{The expressions for frequencies}\label{omegas}
\begin{align}
   & \Omega_\phi^{(0)}=\frac{\sqrt{M}}{a \sqrt{M}+r^{3/2}}\nonumber\\
       & \Omega_\phi^{(1)}=(\Omega_\phi^{(0)})^2 3 (M-r) \sqrt{\frac{M}{r}}\nonumber\\
       & \Omega_\phi^{(2)}=-\frac{3}{2 r^{5/2}} (\Omega_\phi^{(0)})^3 \left(a M^{3/2} \left(29 M^2-26 M r+9 r^2\right)+M r^{3/2} \left(23 M^2-14 M r+3 r^2\right)\right)\nonumber\\
       & \Omega_\phi^{(3)}= (\Omega_\phi^{(0)})^4\frac{6 a   M^{2}}{ r^{3}}  \left(141 M^3-127 M^2 r+51 M r^2-9 r^3\right)\nonumber \\ &
      + (\Omega_\phi^{(0)})^4
        \frac{3a^2  M^{3/2}}{2 r^{9/2}}\left( a^2 M \left(369 M^3-419 M^2 r+207 M r^2-45 r^3\right)+r^3 (3 M-r) \left(71 M^2-24 M r+9 r^2\right)\right)\nonumber\\ &  -  \kappa^2 (\Omega_\phi^{(0)})^2 4 (5 M-3 r) \left(\frac{M}{r}\right)^{9/2}
\end{align}

\begin{align}
   & \Omega_r^{(0)}= \frac{\sqrt{M} \sqrt{-3 a^2+8 a \sqrt{M} \sqrt{r}+r (r-6 M)}}{r \left(a \sqrt{M}+r^{3/2}\right)}\nonumber\\
       & \Omega_r^{(1)}=\frac{(\Omega_\phi^{(0)})^3}{r^4\Omega_r^{(0)}}(3 a^3 M (r-M)+4 a^2 \left(M^2-6 M r+3 r^2\right) \sqrt{M r}\nonumber\\ &+3 a M r (M-r) (2 M+11 r)+20 M^{3/2} r^{7/2}-12 (M r)^{5/2})\nonumber\\
       & \Omega_r^{(2)}=\frac{3M(\Omega_\phi^{(0)})^6}{2r^8(\Omega_r^{(0)})^3}\big(-3 a^6 M \left(109 M^2-66 M r+13 r^2\right)\nonumber\\ &+a^5 \sqrt{M r} \left(1164 M^3-1657 M^2 r+750 M r^2-141 r^3\right)\nonumber\\ &-12 M r^5 \left(24 M^3+8 M^2 r-17 M r^2+4 r^3\right)\nonumber\\ & +a^2 M r^2 \left(612 M^4+1332 M^3 r-4283 M^2 r^2+2026 M r^3-167 r^4\right) \nonumber\\ & -2 a^3 \sqrt{M} r^{3/2} \left(132 M^4+2184 M^3 r-2765 M^2 r^2+1042 M r^3-93 r^4\right) \nonumber\\ & 
       -2 a^4 r \left(554 M^4-2149 M^3 r+1574 M^2 r^2-441 M r^3+24 r^4\right) \nonumber\\ & 
       +a \sqrt{M} r^{7/2} \left(108 M^4+1428 M^3 r-537 M^2 r^2-154 M r^3+39 r^4\right)
       \big) \nonumber\\ & \Omega_r^{(3)}= \frac{M^{3/2}(\Omega_\phi^{(0)})^9}{2r^12(\Omega_r^{(0)})^5}\big(
       -27 a^9 M^{3/2} \left(2077 M^3-1663 M^2 r+583 M r^2-85  r^3 \right) \nonumber\\ & 
      +36 a^8 M \sqrt{r} \left(9137 M^4-12556 M^3 r+6794 M^2 r^2-1914 M r^3+243 r^4\right)
      \nonumber\\ & 
      +16 M^2 r^{15/2} \left(2160 M^4-8784 M^3 r+6609 M^2 r^2-2051 M r^3+258 r^4\right)
          \nonumber\\ & 
          +3 a^7 \sqrt{M} r \left(-238882 M^5+571507 M^4 r-441414 M^3 r^2+172164 M^2 r^3-34416 M r^4+2241 r^5\right)
                    \nonumber\\ & 
                    +a^5 \sqrt{M} r^2 \big(127116 M^6+2564912 M^5 r-5044515 M^4 r^2+3340092 M^3 r^3\nonumber\\&-1094166 M^2 r^4+175140 M r^5-11043 r^6 \big)
                                        \nonumber\\ & 
                                        -4 a^4 M r^{5/2} \big(123876 M^6+73632 M^5 r-871637 M^4 r^2+805428 M^3 r^3\nonumber\\&-302358 M^2 r^4+56394 M r^5-4815 r^6\big)
                                         \nonumber\\ & 
                                         +a \sqrt{M} r^6 \big(196344 M^6-347868 M^5 r+587598 M^4 r^2\nonumber\\&-347353 M^3 r^3+92943 M^2 r^4-10935 M r^5+135 r^6\big)
                                            \nonumber\\ & 
                         +8 a^6 r^{3/2} \big(74252 M^6-388590 M^5 r+443175 M^4 r^2\nonumber\\&-224937 M^3 r^3+59616 M^2 r^4-6912 M r^5+216 r^6\big)                                                          \nonumber\\ & 
       +8 a^2 M r^{9/2} \big(46332 M^6-70200 M^5 r+49869 M^4 r^2\nonumber\\&-82073 M^3 r^3+45021 M^2 r^4-10614 M r^5+1053 r^6\big)
        \nonumber\\ & 
        +a^3 \sqrt{M} r^3 \big(220536 M^7-843948 M^6 r-526902 M^5 r^2+1148387 M^4 r^3\nonumber\\&-288846 M^3 r^4-19392 M^2 r^5+14868 M r^6-2079 r^7\big) \big)
                \nonumber\\ &+\kappa^2 \frac{4M^{9/2}(\Omega_\phi^{(0)})^3}{r^8\Omega_r^{(0)}} \big(3 a^3 \sqrt{M} (13 M-5 r)+2 r^{5/2} \left(18 M^2+5 M r-6 r^2\right)\nonumber\\
               &-6 a^2 \sqrt{r} \left(10 M^2-11 M r+4 r^2\right) +a \sqrt{M} r \left(6 M^2-67 M r+21 r^2\right)\big)
\end{align}

\begin{align}
    &\Omega_\theta^{(0)}=\frac{\sqrt{M} \sqrt{3 a^2-4 a \sqrt{M r}+r^2}}{a \sqrt{M} r+r^{5/2}}\nonumber\\
       &\Omega_\theta^{(1)}=\frac{3 M^2
       }{\left(a \sqrt{M}+r^{3/2}\right)^3 r^4 \Omega_\theta^{(0)}} \big[a^3 \sqrt{M} (r-3 M)+2 a^2 \sqrt{r} (M-r) (M+r)\nonumber\\&-2 a \sqrt{M} r^2 (M-2 r)+r^{7/2} (M-r)\big]\nonumber\\
      & \Omega_{\theta}^{(2)}=\frac{3 M^4}{2 r^8 \left(a \sqrt{M}+r^{3/2}\right)^6(\Omega_\theta^{(0)})^3} \big[3 a^6 M \left(59 M^2-42 M r+11 r^2\right)\nonumber\\ &-a^5 \sqrt{M r} \left(362 M^3-291 M^2 r+12 M r^2+33 r^3\right)\nonumber\\ &+4 a^4 r \left(39 M^4+13 M^3 r-90 M^2 r^2+48 M r^3-3 r^4\right)\nonumber\\ &-2 a^3 \sqrt{M} r^{5/2} \left(121 M^3-269 M^2 r+117 M r^2+3 r^3\right)\nonumber\\ &-4 a^2 r^4 \left(25 M^3+21 M^2 r-24 M r^2+3 r^3\right)\nonumber\\ &+a \sqrt{M} r^{11/2} \left(125 M^2-74 M r+9 r^2\right)\nonumber\\ &+r^7 \left(-23 M^2+14 M r-3 r^2\right)\big]\nonumber\\
      & \Omega_{\theta}^{(3)}=\frac{M^6}{2 r^{12} \left(a \sqrt{M}+r^{3/2}\right)^9(\Omega_\theta^{(0)})^5} \big[-9 a^9 M^{3/2} \left(2253 M^3-2293 M^2 r+999 M r^2-183 r^3\right)\nonumber\\
      &+6 a^8 M \sqrt{r} \left(11505 M^4-14839 M^3 r+6921 M^2 r^2-1197 M r^3-54 r^4\right)\nonumber\\
      &-a^7 \sqrt{M} r \left(74280 M^5-107122 M^4 r+28965 M^3 r^2+13365 M^2 r^3-8289 M r^4+729 r^5\right)\nonumber\\
      &-2 a^6 r^{3/2} \left(-12084 M^6+7039 M^5 r+45718 M^4 r^2-45810 M^3 r^3+15966 M^2 r^4-1809 M r^5+108 r^6\right)\nonumber\\
      &-2 a^5 \sqrt{M} r^3 \left(12522 M^5-65795 M^4 r+56605 M^3 r^2-14388 M^2 r^3-27 M r^4+135 r^5\right)\nonumber\\
      &-2 a^4 r^{9/2} \left(19137 M^5-9680 M^4 r-14234 M^3 r^2+9024 M^2 r^3-1809 M r^4+162 r^5\right)\nonumber\\
      &+2 a^3 \sqrt{M} r^6 \left(9123 M^4-22054 M^3 r+9452 M^2 r^2-1746 M r^3+189 r^4\right)\nonumber\\
      &+a^2 r^{15/2} \left(8337 M^4+3175 M^3 r-3227 M^2 r^2+981 M r^3-162 r^4\right)\nonumber\\
      &+2 a \sqrt{M} r^9 \left(-2541 M^3+1463 M^2 r-459 M r^2+81 r^3\right)+3 r^{21/2} \left(213 M^3-143 M^2 r+51 M r^2-9 r^3\right)\big]\nonumber\\
      &+\kappa^2 \frac{4M^6}{r^{8} \left(a \sqrt{M}+r^{3/2}\right)^3\Omega_\theta^{(0)}}\big[a^3 \sqrt{M} (3 M-5 r)-2 a^2 \sqrt{r} \left(M^2+4 M r-2 r^2\right)\nonumber\\&+2 a \sqrt{M} r^2 (9 M-4 r)+r^{7/2} (3 r-5 M)\big]
\end{align}


\begin{thebibliography}{99}	
		\bibitem{EventHorizonTelescope:2019dse}
		K.~Akiyama \textit{et al.} [Event Horizon Telescope],
		``First M87 Event Horizon Telescope Results. I. The Shadow of the Supermassive Black Hole,''
		Astrophys. J. Lett. \textbf{875}, L1 (2019)
		doi:10.3847/2041-8213/ab0ec7
		[\href{https://arxiv.org/abs/1906.11238}{arXiv:1906.11238 [astro-ph.GA]}].
		
		\bibitem{kerr ds2}
		B. Carter in Les Astres Occlus ed. by B. DeWitt, C. M. DeWitt, (Gordon and Breach, New York, 1973).
\bibitem{Newman:1965tw}
E.~T.~Newman and A.~I.~Janis,
``Note on the Kerr spinning particle metric,''
J. Math. Phys. \textbf{6} (1965), 915-917
doi:10.1063/1.1704350
\bibitem{Reissner}
H.~ Reissner, "Über die Eigengravitation des elektrischen Feldes nach der Einsteinschen Theorie". Annalen der Physik (in German). 50 (9): 106–120,  (1916).
\bibitem{Nordstrom}
 G.~Nordström,  "On the Energy of the Gravitational Field in Einstein's Theory". Verhandl. Koninkl. Ned. Akad. Wetenschap., Afdel. Natuurk., Amsterdam. 26: 1201–1208, (1918).
 \bibitem{Ovalle:2021jzf}
		J.~Ovalle, E.~Contreras and Z.~Stuchlik,
		Kerr\textendash{}de Sitter black hole revisited,
		Phys. Rev. D \textbf{103}, no.8, 084016 (2021)
		doi:10.1103/PhysRevD.103.084016
		[\href{https://arxiv.org/abs/2104.06359}{arXiv:2104.06359 [gr-qc]}].
\bibitem{Brahma:2020eos}
S.~Brahma, C.~Y.~Chen and D.~h.~Yeom,
``Testing Loop Quantum Gravity from Observational Consequences of Nonsingular Rotating Black Holes,''
Phys. Rev. Lett. \textbf{126} (2021) no.18, 181301
doi:10.1103/PhysRevLett.126.181301
[arXiv:2012.08785 [gr-qc]].
\bibitem{Abedi:2015yga}
J.~Abedi and H.~Arfaei,
``Obstruction of black hole singularity by quantum field theory effects,''
JHEP \textbf{03} (2016), 135
doi:10.1007/JHEP03(2016)135
[arXiv:1506.05844 [gr-qc]].
\bibitem{Bojowald:2015zha}
M.~Bojowald, S.~Brahma and J.~D.~Reyes,
``Covariance in models of loop quantum gravity: Spherical symmetry,''
Phys. Rev. D \textbf{92} (2015) no.4, 045043
doi:10.1103/PhysRevD.92.045043
[arXiv:1507.00329 [gr-qc]].
\bibitem{Bojowald:2018xxu}
M.~Bojowald, S.~Brahma and D.~h.~Yeom,
``Effective line elements and black-hole models in canonical loop quantum gravity,''
Phys. Rev. D \textbf{98} (2018) no.4, 046015
doi:10.1103/PhysRevD.98.046015
[arXiv:1803.01119 [gr-qc]].
\bibitem{Bojowald:2020unm}
M.~Bojowald,
``No-go result for covariance in models of loop quantum gravity,''
Phys. Rev. D \textbf{102} (2020) no.4, 046006
doi:10.1103/PhysRevD.102.046006
[arXiv:2007.16066 [gr-qc]].
\bibitem{Yang:2023cmv}
H.~Yang, C.~J.~Yu and Y.~G.~Miao,
``Preliminary analyses on dynamics and thermodynamics of rotating regular black holes,''
[arXiv:2308.03068 [gr-qc]].
\bibitem{Tu:2023xab}
Z.~Y.~Tu, T.~Zhu and A.~Wang,
Phys. Rev. D \textbf{108} (2023) no.2, 2
doi:10.1103/PhysRevD.108.024035
[arXiv:2304.14160 [gr-qc]].
\bibitem{Liu:2023vfh}
C.~Liu, H.~Xu, H.~Siew, T.~Zhu, Q.~Wu and Y.~Zhao,
``Constraints on the rotating self-dual black hole with quasi-periodic oscillations,''
[arXiv:2305.12323 [gr-qc]].
\bibitem{Jiang:2023img}
H.~X.~Jiang, C.~Liu, I.~K.~Dihingia, Y.~Mizuno, H.~Xu, T.~Zhu and Q.~Wu,
``Shadows of Loop Quantum Black Holes: Semi-analytical Simulations of Loop Quantum Gravity Effects on Sagittarius A* and M 87*,''
[arXiv:2312.04288 [gr-qc]].

\bibitem{Belloni:2009ph}
T.~M.~Belloni,
Lect. Notes Phys. \textbf{794} (2010), 53-84
doi:10.1007/978-3-540-76937-8\_3
[arXiv:0909.2474 [astro-ph.HE]].
\bibitem{Abramowicz:2001bi}
M.~A.~Abramowicz and W.~Kluzniak,
Astron. Astrophys. \textbf{374} (2001), L19
doi:10.1051/0004-6361:20010791
[arXiv:astro-ph/0105077 [astro-ph]].
\bibitem{Pasham:2014ybe}
D.~R.~Pasham, T.~E.~Strohmayer and R.~F.~Mushotzky,
Nature \textbf{513} (2014), 74
doi:10.1038/nature13710
[arXiv:1501.03180 [astro-ph.HE]].
\bibitem{alireza}
A.~Allahyari and L.~Shao,
JCAP \textbf{10} (2021), 003
doi:10.1088/1475-7516/2021/10/003
[arXiv:2102.02232 [gr-qc]].
\bibitem{samimi}
SAMIMI, J., SHARE, G., WOOD, K. et al. GX339–4: a new black hole candidate. Nature 278, 434–436 (1979). https://doi.org/10.1038/278434a0

\bibitem{Haggard:2014rza}
H.~M.~Haggard and C.~Rovelli,
``Quantum-gravity effects outside the horizon spark black to white hole tunneling,''
Phys. Rev. D \textbf{92} (2015) no.10, 104020
doi:10.1103/PhysRevD.92.104020
[arXiv:1407.0989 [gr-qc]].
\bibitem{Bianchi:2018mml}
E.~Bianchi, M.~Christodoulou, F.~D'Ambrosio, H.~M.~Haggard and C.~Rovelli,
``White Holes as Remnants: A Surprising Scenario for the End of a Black Hole,''
Class. Quant. Grav. \textbf{35} (2018) no.22, 225003
doi:10.1088/1361-6382/aae550
[arXiv:1802.04264 [gr-qc]].
\bibitem{Olmedo:2017lvt}
J.~Olmedo, S.~Saini and P.~Singh,
``From black holes to white holes: a quantum gravitational, symmetric bounce,''
Class. Quant. Grav. \textbf{34} (2017) no.22, 225011
doi:10.1088/1361-6382/aa8da8
[arXiv:1707.07333 [gr-qc]].
\bibitem{BenAchour:2020gon}
J.~Ben Achour, S.~Brahma, S.~Mukohyama and J.~P.~Uzan,
``Towards consistent black-to-white hole bounces from matter collapse,''
JCAP \textbf{09} (2020), 020
doi:10.1088/1475-7516/2020/09/020
[arXiv:2004.12977 [gr-qc]].
\bibitem{Martin-Dussaud:2019wqc}
P.~Martin-Dussaud and C.~Rovelli,``Evaporating black-to-white hole,''
Class. Quant. Grav. \textbf{36} (2019) no.24, 245002
doi:10.1088/1361-6382/ab5097
[arXiv:1905.07251 [gr-qc]].
\bibitem{Rignon-Bret:2021jch}
A.~Rignon-Bret and C.~Rovelli,
``Black to white transition of a charged black hole,''
Phys. Rev. D \textbf{105} (2022) no.8, 086003
doi:10.1103/PhysRevD.105.086003
[arXiv:2108.12823 [gr-qc]].
\bibitem{Han:2023wxg}
M.~Han, C.~Rovelli and F.~Soltani,
Phys. Rev. D \textbf{107} (2023) no.6, 064011
doi:10.1103/PhysRevD.107.064011
[arXiv:2302.03872 [gr-qc]].







\bibitem{Bodendorfer:2019jay}
N.~Bodendorfer, F.~M.~Mele and J.~M\"unch,
``Mass and Horizon Dirac Observables in Effective Models of Quantum Black-to-White Hole Transition,''
Class. Quant. Grav. \textbf{38} (2021) no.9, 095002
doi:10.1088/1361-6382/abe05d
[arXiv:1912.00774 [gr-qc]].
\bibitem{Hong:2022thd}
D.~K.~Hong, W.~C.~Lin and D.~h.~Yeom,
``Trouble with geodesics in black-to-white hole bouncing scenarios,''
Phys. Rev. D \textbf{106} (2022) no.10, 104011
doi:10.1103/PhysRevD.106.104011
[arXiv:2207.03183 [gr-qc]].
\bibitem{Jalalzadeh:2022rxx}
S.~Jalalzadeh,
``Quantum black hole\textendash{}white hole entangled states,''
Phys. Lett. B \textbf{829} (2022), 137058
doi:10.1016/j.physletb.2022.137058
[arXiv:2203.09968 [gr-qc]].







\bibitem{Bodendorfer:2019nvy}
N.~Bodendorfer, F.~M.~Mele and J.~M\"unch,
``(b,v)-type variables for black to white hole transitions in effective loop quantum gravity,''
Phys. Lett. B \textbf{819} (2021), 136390
doi:10.1016/j.physletb.2021.136390
\bibitem{Bardeen:1972fi}
J.~M.~Bardeen, W.~H.~Press and S.~A.~Teukolsky,
``Rotating black holes: Locally nonrotating frames, energy extraction, and scalar synchrotron radiation,''
Astrophys. J. \textbf{178} (1972), 347
doi:10.1086/151796
\bibitem{Carroll:1997ar}
S.~M.~Carroll,
``Lecture notes on general relativity,''
[arXiv:gr-qc/9712019 [gr-qc]].
\bibitem{Suzuki:1997by}
S.~Suzuki and K.~i.~Maeda,
Phys. Rev. D \textbf{58} (1998), 023005
doi:10.1103/PhysRevD.58.023005
[arXiv:gr-qc/9712095 [gr-qc]].
\bibitem{Zhang:2017nhl}
Y.~P.~Zhang, S.~W.~Wei, W.~D.~Guo, T.~T.~Sui and Y.~X.~Liu,
``Innermost stable circular orbit of spinning particle in charged spinning black hole background,''
Phys. Rev. D \textbf{97} (2018) no.8, 084056
doi:10.1103/PhysRevD.97.084056
[arXiv:1711.09361 [gr-qc]].

\bibitem{Jefremov:2015gza}
P.~I.~Jefremov, O.~Y.~Tsupko and G.~S.~Bisnovatyi-Kogan,
``Innermost stable circular orbits of spinning test particles in Schwarzschild and Kerr space-times,''
Phys. Rev. D \textbf{91} (2015) no.12, 124030
doi:10.1103/PhysRevD.91.124030
[arXiv:1503.07060 [gr-qc]].

\bibitem{Bambi:2013fea}
C.~Bambi,
``Testing the nature of the black hole candidate in GRO J1655-40 with the relativistic precession model,''
Eur. Phys. J. C \textbf{75} (2015) no.4, 162
doi:10.1140/epjc/s10052-015-3396-7
[arXiv:1312.2228 [gr-qc]].
\bibitem{Kaplan} S. A. Kaplan, JETP, 19, 951 (1949)
\bibitem{Landau} L. D. Landau, E. M. Lifshitz, The Classical Theory of
Fields. Pergamon, Oxford (1993)
\bibitem{Ruffini}
R. Ruffini, J. Wheeler, Cosmology from space platform in
Proceedings of the Conference on Space Physics. – Paris:
ESRO (1971)

\bibitem{Colistete:2002ka}
R.~Colistete, Jr., C.~Leygnac and R.~Kerner,
``Higher order geodesic deviations applied to the Kerr metric,''
Class. Quant. Grav. \textbf{19} (2002), 4573-4590
doi:10.1088/0264-9381/19/17/309
[arXiv:gr-qc/0205019 [gr-qc]].
\bibitem{Motta:2013wga}
S.~E.~Motta, T.~M.~Belloni, L.~Stella, T.~Mu\~noz-Darias and R.~Fender,
``Precise mass and spin measurements for a stellar-mass black hole through X-ray timing: the case of GRO J1655\ensuremath{-}40,''
Mon. Not. Roy. Astron. Soc. \textbf{437} (2014) no.3, 2554-2565
doi:10.1093/mnras/stt2068
[arXiv:1309.3652 [astro-ph.HE]].
\bibitem{Belloni:2016xgi}
T.~M.~Belloni and S.~E.~Motta,
``Transient Black Hole Binaries,''
doi:10.1007/978-3-662-52859-4\_2
[arXiv:1603.07872 [astro-ph.HE]].


\bibitem{Ashtekar:2021kfp}
A.~Ashtekar and E.~Bianchi,
Rept. Prog. Phys. \textbf{84}, no.4, 042001 (2021)
doi:10.1088/1361-6633/abed91
[arXiv:2104.04394 [gr-qc]].

\end{thebibliography}
\end{document}